\documentclass[twocolumn,showpacs,preprintnumbers,amsmath,amssymb]{revtex4}
\usepackage{graphicx}% Include figure files
\usepackage{dcolumn}% Align table columns on decimal point
\usepackage{bm}% bold math

\newcommand{\mean}[1]{\langle{#1}\rangle}
\newcommand{\pro}[2]{\langle{#1}|{#2}\rangle}
\newcommand{\bra}[1]{\langle{#1}|}
\newcommand{\ket}[1]{|{#1}\rangle}

\newcommand{\dgg}{^{\dagger}}

\begin{document}
\title{Dynamical Gaussian state transfer with quantum error correcting 
architecture}%

\author{Go Tajimi}%
\email[E-mail address: ]{go@z8.keio.jp}
\author{Naoki Yamamoto}
\email[E-mail address: ]{yamamoto@appi.keio.ac.jp}
\affiliation{Department of Applied Physics and Physico-Informatics, 
Keio University, Yokohama 223-8522, Japan}
\date{\today}%

\begin{abstract}

Transferring a quantum state of a light field to a memory is of 
particular importance. 
However, this transfer is usually hampered because the memory 
system is subjected to some noise and this can limit the performance 
of the state transfer to a great extent. 
In this paper, we consider the transfer of a Gaussian state of 
light to a linear medium memory such as an 
opto-mechanical oscillator and propose a dynamical feedback 
controller that suppresses the noise in the memory system. 
To protect an unknown state, the feedback scheme employs the 
specific configuration of the quantum error correction; 
that is, a three-mode Gaussian state having appropriate syndromes 
is taken as the input. 
Correspondingly, the memory consists of three independent 
linear systems. 
The syndrome errors are estimated continuously in time through 
the measurement of the output field, and the results are 
then fed back to control the system. 
Because the input is Gaussian and the systems are all linear, 
it is possible to formulate the problem using the framework 
of the celebrated classical Kalman filtering and linear 
quadratic Gaussian control. 
A numerical simulation demonstrates the effectiveness of the 
control scheme. 

\end{abstract}


\maketitle
% \tableofcontents

%%%%%%%%%%%%%%%%%%%%%%%%%%%%%%%%%%%%%%%%%%%%%%%%%%%%%%%%%%%%%%%%
%%%%%%%%%%%%%%%%%%%%%%%%%%%%%%%%%%%%%%%%%%%%%%%%%%%%%%%%%%%%%%%%
%%%%%%%%%%%%%%%%%%%%%%%%%%%%%%%%%%%%%%%%%%%%%%%%%%%%%%%%%%%%%%%%

\section{Introduction}

Transferring a quantum state of light to a memory is 
of particular importance for various purposes in quantum 
information technologies 
\cite{Maitre1997,Fleischhauer2002,Giovanetti2005,Meekhof1996,
Cirac1997,Phillips2001,Schori2002,Dantan2004,Paternostro2005,
Burgarth2005,Paris2009,Paris2010,Kozhekin2000,Julsgaard2001,
Julsgaard2004,matsukevich2004quantum,Appel2008,He2009,
Parkins1999,Zhang2003,Filip2009,Filip2010}. 
Candidates for a memory are largely divided into two categories: 
discrete variable systems such as an atom with distinct 
energy levels 
\cite{Maitre1997,Fleischhauer2002,Giovanetti2005,Meekhof1996,
Cirac1997,Phillips2001,Schori2002,Dantan2004,
matsukevich2004quantum,Paternostro2005,Burgarth2005,Paris2009,
Paris2010} and continuous variable systems such as an 
opto-mechanical oscillator with a vibration mode 
\cite{Kozhekin2000,Julsgaard2001,Julsgaard2004,Appel2008,He2009,
Parkins1999,Zhang2003,Filip2009,Filip2010}. 
Remarkably, some experimental demonstrations of quantum state 
transfer have been reported 
\cite{Meekhof1996,Julsgaard2001,Julsgaard2004,
matsukevich2004quantum,Appel2008}.

In reality, however, the memory performance is essentially 
limited by environmental noise that inevitably occurs during 
the transfer process. 
In a real experiment several state-of-the-art techniques should 
be, at least implicitly, employed to suppress such noise. 
However, to the best of our knowledge there exists no systematic 
method for suppressing the noise. 
Therefore, control theory for quantum memory needs to be explored.

In working out this subject, a key fact is that due to 
the dynamical noise the information contained in the memory 
state is gradually erased, rather than that the noise suddenly 
vanishes it. 
Hence, if such an unwanted change of state can be monitored 
continuously in time, then the measurement result could be fed 
back to suppress the noise. 
In fact, in \cite{Cirac1997}, the probe light field for state 
transfer is continuously measured at the terminal of the optical 
path to indirectly detect the error during the transfer process. 
In this paper, we propose a dynamical control scheme that uses such 
continuous measurement results not only for the error 
detection but further for feedback control to obtain the better 
performance of the state transfer. 
Fortunately, measurement-based quantum feedback 
control theory is well developed \cite{Belavkin1993,Bouten2009,
Wiseman2009book} and has many practical uses 
\cite{Doherty1999,Doherty2000,Thomsen2002,Stockton2004,
Edwards2005,Yamamoto2006,Mirrahimi2007}; 
hence, it will be also useful for our scheme.

Here we describe two specific features of the proposed 
system-controller configuration. 
The first one is that the memory is a linear medium system 
such as a collective atomic ensemble or an opto-mechanical 
oscillator. 
In addition, the input state to be transferred is assumed to 
be a Gaussian state of light \cite{ferraro2005}, in which case 
the memory state also becomes Gaussian. 
Then, the feedback control scheme has the form of celebrated 
{\it linear quadratic Gaussian (LQG)} control with {\it Kalman 
filtering} \cite{Doherty1999,Stockton2004,Edwards2005,
Yamamoto2006,Wiseman2009book}. 
The second feature is that the controller employes the schematic 
of quantum error correction (QEC) for the purpose of protecting 
an {\it unknown} input state. 
The basic idea of QEC is to encode an unknown state into an 
enlarged codespace so that an appropriate {\it syndrome} measurement 
yields sufficient information about the error, which can then 
be used to correct the error 
\cite{Shor1995,Lloyd1998,Braunstein1998prl,Braunstein1998nature,
van2008note,Aoki2009}. 
Use of this QEC strategy for quantum memory means that the 
input is a three-mode unknown Gaussian state having appropriate 
syndromes, and it is transferred to the memory that is 
correspondingly enlarged; 
during the transfer process, the output light fields are 
measured continuously in time, yielding the estimate of the 
syndrome errors that can be fed back to control the system.

The effectiveness of this control scheme is actually expected 
from the fact that the dynamical feedback control has successfully 
been applied to some QEC problems \cite{Ahn2002,Ahn2003,Sarovar2004,
Oreshkov2007,Chase2008,Mabuchi2009,Kerckhoff2010}. 
It should be noted that the problem considered in this paper 
is not a QEC problem itself; 
in QEC, the system state can be protected with the use of ancillary 
apparatus, whereas in our formalism, the initial state of the 
memory system cannot be protected, but rather the input state to 
be transferred to the system is what should be protected. 
This observation is vital in the sense that our scheme 
does not violate the no-go theorem for Gaussian QEC 
\cite{niset2009}.

The paper is organized as follows. 
In Section II, as a preliminary, we describe a single-mode Gaussian 
state transfer. 
The dynamical control scheme is addressed in two sections. 
First, the state preparation and transfer processes are given 
in Section III. 
Section IV explains the error detection scheme (Kalman filter) 
and the feedback controller (LQG controller). 
Finally, in Section V, we numerically demonstrate the efficiency 
of our method by taking, as an example of the memory, an 
opto-mechanical oscillator manipulated under ultra-low 
temperature.

%%%%%%%%%%%%%%%%%%%%%%%%%%%%%%%%%%%%%%%%%%%%%%%%%%%%%%%%%%%%%%%%
%%%%%%%%%%%%%%%%%%%%%%%%%%%%%%%%%%%%%%%%%%%%%%%%%%%%%%%%%%%%%%%%
%%%%%%%%%%%%%%%%%%%%%%%%%%%%%%%%%%%%%%%%%%%%%%%%%%%%%%%%%%%%%%%%

\section{Single-mode Gaussian state transfer}

\begin{figure}[!htbp]
\includegraphics[width=50mm]{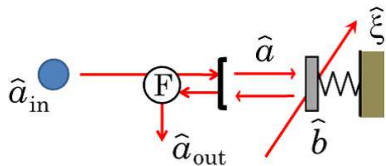}
\caption{
(Color online) 
Opto-mechanical oscillator as an example of the single-mode 
linear quantum memory. 
The right end-mirror of the Fabry-Perot cavity serves as 
the oscillator, where $\hat a$ and $\hat b$ are annihilation 
operators corresponding to the intra-cavity and oscillator 
modes, respectively. 
Adiabatically eliminating the cavity mode $\hat a$, 
Eq. \eqref{b-Langevin} is obtained. 
$F$ denotes a Faraday isolator that facilitates one-way 
coupling between the input and cavity modes. 
}
\label{fig1}
\end{figure}

As a preliminary, this section is devoted to describe a 
single-mode Gaussian state transfer from a light field 
to a memory, where the former is a coherent CW laser field 
and the latter is served by a linear system such as an 
opto-mechanical oscillator shown in Fig.~\ref{fig1} 
(see e.g., \cite{Parkins1999,Zhang2003,law1995} 
for a detailed discussion). 
The input is a coherent state $\ket{\alpha_{\rm in}}$, with 
$\hat a_{\rm in}$ the annihilation operator of this 
coherent light field. 
The reflected output mode $\hat a_{\rm out}$ could be used 
to correct the errors occurring during the transfer process. 
In what follows we describe the dynamics of the memory system 
and show how much the input state $\ket{\alpha_{\rm in}}$ is 
transferred to the memory.

A general single-mode open linear dynamics 
is described by the following quantum Langevin equation:
\begin{align}
\label{b-Langevin}
    \frac{d\hat b}{dt}
       =-\frac{\nu+\Gamma}{2}\hat b
            -\sqrt{\nu}( \alpha_{\rm in} + \hat a_0)
			    - \sqrt{\Gamma}\hat \xi, 
\end{align}
where $\hat b$ denotes the system annihilation operator. 
Here, $\nu$ represents the coupling constant between 
the system and the input coherent light field 
$\hat a_{\rm in}=\alpha_{\rm in}+\hat a_0$, where $\alpha_{\rm in}$ 
is the mean and $\hat a_0$ is the field annihilation operator. 
Moreover, we assume that the system couples with an unwanted 
single-mode noisy environment represented by the 
annihilation operator $\hat \xi$ with mean zero. 
$\Gamma$ is the coupling strength.

Let us now take the white noise approximation on the outer field 
modes $\hat a_0$ and $\hat \xi$; 
i.e., 
$\langle \hat a_0(t)\hat a_0^\dagger(t')\rangle = \delta(t-t')$ and 
$\langle \hat \xi(t)\hat \xi^\dagger(t')\rangle = (n+1)\delta(t-t')$, 
where $n>0$ represents the strength of the noise. 
In the case of thermal noise, $n$ is the averaged photon number. 
This approximation allows us to represent the dynamics of $\hat b$ 
in terms of the Ito-type quantum stochastic differential equation 
(QSDE) \cite{hudson1984quantum,Gardiner}:
\begin{align}
\label{b-Ito}
    d\hat b = -\frac{\nu+\Gamma}{2}\hat b dt
                  -\sqrt{\nu}\alpha _{\rm in}dt-\sqrt{\nu}d\hat A_0
                         -\sqrt{\Gamma}d\hat \Xi, 
\end{align}
where $\hat A_0(t)=\int_0^t \hat a_0(s)ds$ and 
$\hat \Xi(t)=\int_0^t \hat \xi(s) ds$ are the so-called quantum 
Wiener processes. 
Their infinitesimal increments satisfy the following quantum 
Ito-rule:
\begin{align*}
     &
     d\hat A_0 d\hat A_0^{\dag} = dt,~
     (d\hat A_0)^2=d\hat A_0^{\dag 2}
                       =d\hat A_0^{\dag} d\hat A_0=0, 
\nonumber \\
     &
     d\hat \Xi d\hat \Xi^\dag = (n+1)dt,~
     d\hat \Xi^\dag d\hat \Xi = n dt,~
     d\hat \Xi^2=d\hat \Xi^{\dag 2}=0. 
\end{align*}
In this QSDE representation the input mode is written by 
$d\hat A_{\rm in}=\alpha_{\rm in}dt+d\hat A_0$. 
Since the field state is the vacuum, we have 
$\langle d\hat A_{\rm in}\rangle =\alpha _{\rm in}dt$. 
Also note that $\mean{d\hat \Xi}=0$.

Now, let us see the steady state of the memory system. 
Using the above Ito rule we readily obtain the ordinary 
differential equations of $\mean{b(t)}$ and $\mean{b(t)\dgg b(t)}$, 
which give 
$\mean{\hat b(\infty)} = -2\sqrt{\nu}\alpha_{\rm in}/(\nu+\Gamma)$ 
and 
\[
   \mean{\Delta \hat q(\infty)^2} 
     = \mean{\Delta \hat p(\infty)^2} 
       = \frac{1}{2} + \frac{\Gamma n}{\nu +\Gamma} 
       > \frac{1}{2}, 
\]
where $\hat q=(\hat b+\hat b^\dag )/\sqrt{2}$ and 
$\hat p=(\hat b-\hat b^\dag )/\sqrt{2}i$ denote the dimensionless 
position and momentum operators of the system, respectively. 
This clearly illustrates that in the long time limit the system 
certainly acquires the information of the field input state 
$\ket{\alpha_{\rm in}}$ in the mean sense (note that $\nu$ and 
$\Gamma$ are assumed to be known). 
However the fluctuation of the state could become much bigger than 
the vacuum fluctuation $1/2$; 
in this case the state wrote down to the memory is far away 
from pure and has very small overlap with the input state.

%%%%%%%%%%%%%%%%%%%%%%%%%%%%%%%%%%%%%%%%%%%%%%%%%%%%%%%%%%%%%%%%
%%%%%%%%%%%%%%%%%%%%%%%%%%%%%%%%%%%%%%%%%%%%%%%%%%%%%%%%%%%%%%%%
%%%%%%%%%%%%%%%%%%%%%%%%%%%%%%%%%%%%%%%%%%%%%%%%%%%%%%%%%%%%%%%%

\section{Preparation and transfer processes of the input state}

\begin{figure}[htbp]
\begin{center}
\includegraphics[width=84mm]{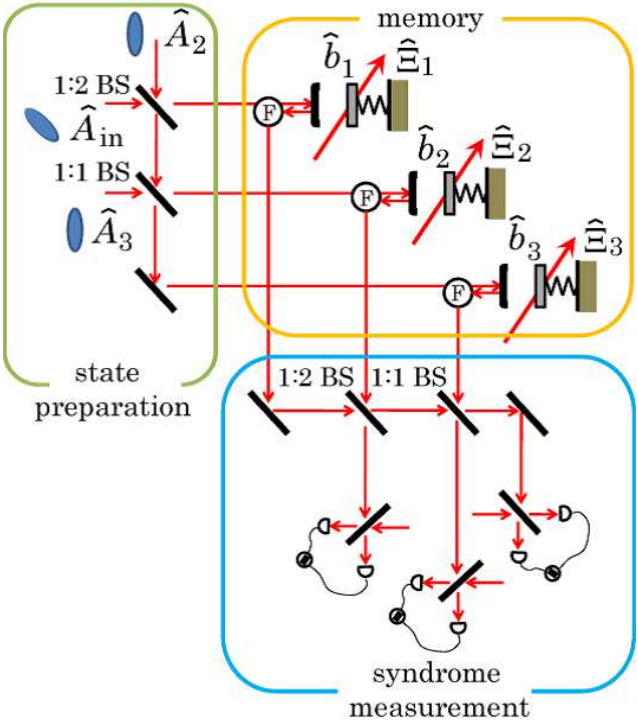}
\caption{
(Color online) 
A schematic of the dynamical state transfer with QEC architecture, 
where in this picture an opto-mechanical oscillator is taken as 
a memory system. 
}
\label{physical setup}
\end{center}
\end{figure}

In this paper, we study the system shown in 
Fig.~\ref{physical setup}. 
First, in the left box, the input state, which is a three-mode 
Gaussian state of light, is prepared. 
This input state is transferred into the memory shown in the 
middle box in the figure, that correspondingly consists of 
three identical linear systems; 
in the figure, opto-mechanical oscillators are particularly 
depicted as the memory system. 
Finally, the output fields are measured, and the results 
are used for feedback control.

A notable feature of this transfer scheme is that the input 
state is allowed to contain some unknown parameters. 
More specifically, the input state is generated by embedding 
an unknown single-mode field $\hat A_{\rm in}$ into the 
three-mode light fields with the use of ancilla squeezed 
fields and beam splitters. 
This is no more than the schematic of QEC; 
actually, $\hat A_{\rm in}$ corresponds to a {\it source mode} 
to be protected, and for this purpose it is encoded into a larger 
Hilbert space where we can construct appropriate syndromes 
that do not depend on the unknown source mode but only contain 
information about the errors. 
Those errors could be corrected by feedback control through 
the syndrome measurement depicted in the lower box of the 
figure. 
Note that the three-mode encoding allows us to detect only 
either the position or the momentum error, but the proposed 
scheme can be extended to a nine-mode code to detect errors 
acting on both the position and momentum operators 
\cite{Aoki2009}.

In this section, we describe the above-mentioned state 
preparation and transfer processes, and then evaluate 
in detail the memory performance, when without any error 
correction.

%%%%%%%%%%%%%%%%%%%%%%%%%%%%%%%%%%%%%%%%%%%%%%%%%%%%%%%%%%%%%%%%

\subsection{State preparation}

For discrete variable systems, a series of CNOT gates is 
often used for encoding. 
A possible continuous-variable analogue is realized in 
linear optics circuit shown in the left box of 
Fig. \ref{physical setup}; 
first, the information source mode $\hat A_{\rm in}$ is 
mixed via a $1:2$ beam splitter with an ancilla squeezed 
vacuum field denoted by $\hat A_2$, and one of the outputs 
is further combined with the second ancilla squeezed vacuum 
field $\hat A_3$. 
This combination of beam splitters is called the {\it tritter} 
\cite{Aoki2009}.

Let us describe the above encoding process in detail. 
As in the single-mode case discussed in Sec.~II, we write 
the source mode as $d\hat A_{\rm in}=\alpha_{\rm in}dt+d\hat A_1$, 
where $\alpha_{\rm in}$ and $\hat A_1$ are the mean amplitude 
and the quantum fluctuation, respectively. 
In particular, we assume 
\[
     \alpha_{\rm in}\in{\mathbb R}, 
\]
implying that the momentum element of the mean is known to be zero. 
As mentioned above, in this case, the three-mode encoding is 
sufficient to detect the position errors. 
The quantum fluctuation $\hat A_j~(j=1, 2, 3)$ is a stochastic 
process satisfying $\langle{d\hat A_j}\rangle =0$ and 
the following {\it quantum Ito rule}: 
\begin{align*}
     &d\hat A_jd\hat A_j^\dag =(N_j+1)dt,~~~
     d\hat A_j^\dag d\hat A_j =N_jdt, 
\nonumber \\
     &d\hat A_j^2=M_jdt,~~~
     d\hat A_j^{\dag 2}=M_j^\ast dt,
\end{align*}
where the parameters $N_j\geq 0$ and $M_j\in{\mathbb C}$ have to 
satisfy $N_j(N_j+1)\geq |M_j|^2$. 
The fluctuation parameters $N_1$ and $M_1$ of the source mode 
are unknown. 
For the ancilla modes $\hat A_2$ and $\hat A_3$, we set 
\[
    M_2=M_3=\frac{e^\mu-e^{-\mu}}{4},~~
    N_2=N_3=\frac{e^\mu+e^{-\mu}-2}{4},
\]
which satisfy $N_j(N_j+1)=M_j^2$; 
that is, the ancilla modes are the identical pure squeezed 
vacuum fields with squeezing parameter $\mu$. 
Let us collect the field quadratures in a single vector as 
\begin{equation}
\label{noise 1}
    \hat W_1=
      (\hat Q_1, \hat P_1, 
          \hat Q_2, \hat P_2, 
             \hat Q_3, \hat P_3)^\top, 
\end{equation}
where $\hat Q_j=(\hat A_j+\hat A_j^\dag )/\sqrt{2}$ and 
$\hat P_j=(\hat A_j-\hat A_j^\dag )/\sqrt{2}i$. 
Then, through the tritter, $\hat W_1$ becomes 
$d\hat W_1'=\beta dt+Td\hat W_1$, where $T$ is the orthogonal 
matrix corresponding to the combination of the above-mentioned 
$1:2$ and half beam splitters: 
\begin{align}
   &T=
\nonumber \\
   & \begin{pmatrix}
           \sqrt{1/3} & 0& -\sqrt{2/3} & 0 & 0 & 0 \\
           0 & \sqrt{1/3} & 0 & -\sqrt{2/3} & 0 & 0 \\
           \sqrt{1/3} & 0 & \sqrt{1/6} & 0 & \sqrt{1/2} & 0 \\
           0 & \sqrt{1/3} & 0 & \sqrt{1/6} & 0 & \sqrt{1/2} \\
           \sqrt{1/3} & 0 & \sqrt{1/6} & 0 & -\sqrt{1/2} & 0 \\
           0 & \sqrt{1/3} & 0 & \sqrt{1/6} & 0 & -\sqrt{1/2} \\
     \end{pmatrix},
\nonumber
\end{align}
and
\[
   \beta
     =\frac{\sqrt{2}\alpha_{\rm in}}{\sqrt{3}}
        (1, 0, 1, 0, 1, 0)^\top. 
\]
Using the Ito rule, it is found that the position quadratures 
of $\hat W_1'$ satisfy 
\begin{equation}
\label{encoded q-difference}
    \frac{d}{dt}\mean{ (\hat Q'_i - \hat Q'_j)^2}
     =e^\mu, ~~~\forall i\not =j. 
\end{equation}
This quantity is essentially equivalent to the power spectrum 
density of $\mean{ (\hat Q'_i - \hat Q'_j)^2}$ at the center 
frequency of the career laser field \cite{bachor2004guide}. 
We then find that Eq. \eqref{encoded q-difference} becomes zero 
for every $i,j$ when taking the limit $\mu \rightarrow -\infty$. 
This means that, in the Schrodinger picture, this ideal input 
state generated through the tritter lives in the codespace 
spanned by $\{\ket{Q,Q,Q}\}$ and is of the following GHZ-like form : 
\begin{equation}
\label{GHZ}
    \ket{\tilde{\psi}_{\rm in}}=\int\psi(Q)\ket{Q,Q,Q}dQ, 
\end{equation}
where $\psi(Q)$ is a Gaussian wave function of the source 
state. 
The tilde means that the state is the ideal one.

Next let us focus on the following quantity: 
\begin{eqnarray}
& & \hspace*{-2em}
\label{field_entanglement}
    {\cal P}_{\rm fd} :=
        \mean{ (\hat Q'_1 - \hat Q'_2)^2}
          + \mean{ (\hat Q'_2 - \hat Q'_3)^2}
            + \mean{ (\hat Q'_3 - \hat Q'_1)^2} 
\nonumber \\ & & \hspace*{2.5em}
    \mbox{}
          + 3\mean{ (\hat P'_1 + \hat P'_2 + \hat P'_3)^2}, 
\end{eqnarray}
which leads to 
\[
   \frac{d{\cal P}_{\rm fd}}{dt} 
     = 3e^\mu
        +\frac{9}{2}(2N_1-M_1-M_1^\ast+1). 
\]
If $\hat A_1, \hat A_2$, and $\hat A_3$ are all coherent 
fields, we have $d{\cal P}_{\rm fd}/dt=7.5$, which thus can 
be interpreted as the classical limit. 
This means that a non-classical input state is generated when 
$d{\cal P}_{\rm fd}/dt<7.5$, which is now equivalent to that 
$\mu<0$ and $M_1\geq 0$. 
Moreover, it is known that a symmetric state such as the one 
taken here is entangled if $d{\cal P}_{\rm fd}/dt<6$ 
\cite{van2003detecting,simon2000}. 
For instance setting a coherent source state, i.e., $N_1=M_1=0$, 
and taking the ideal limit $\mu \rightarrow -\infty$, we obtain 
$d{\cal P}_{\rm fd}/dt\rightarrow 4.5$; 
hence in this case the GHZ-like state \eqref{GHZ} is 
indeed entangled.

%%%%%%%%%%%%%%%%%%%%%%%%%%%%%%%%%%%%%%%%%%%%%%%%%%%%%%%%%%%%%%%%

\subsection{Transfer process}

Let us next describe the transfer process of the mode $\hat W_1'$ 
 to the memory served by the three identical linear systems. 
The dynamics of the memory is given by the combination of 
Eq. \eqref{b-Ito} with the input field replaced by $d\hat W_1'$; 
that is, the vector of quadratures, 
\[
   \hat x = (\hat q_1, \hat p_1, \hat q_2, \hat p_2, 
                                 \hat q_3, \hat p_3)^\top
\]
with $\hat q_j=(\hat b_j+\hat b_j^\dag )/\sqrt{2}$ and 
$\hat p_j=(\hat b_j-\hat b_j^\dag )/\sqrt{2}i$, satisfies the 
following QSDE:
\begin{eqnarray}
\label{dynamics}
& & \hspace*{-2em}
    d\hat x = A\hat xdt + udt 
                 - \sqrt{\nu}d\hat W_1' 
                    - \sqrt{\Gamma}d\hat W_2
\nonumber \\ & & \hspace*{-0.65em}
    = A\hat xdt + udt - \sqrt{\nu}\beta dt + B d\hat W,
\end{eqnarray}
where $\hat W=(\hat W_1,~\hat W_2)^\top$ and 
\[
    A =-\frac{\nu +\Gamma }{2}I_6,~~~
    B =\begin{pmatrix}-\sqrt {\nu }T,-\sqrt {\Gamma }I_6\end{pmatrix}.
\]
The fluctuation vector $\hat W_1$ is given by Eq. \eqref{noise 1} and 
\[
    \hat W_2 
     = \sqrt{2}(\Re (\hat \Xi_1), \Im (\hat \Xi_1), 
                  \Re (\hat \Xi_2), \Im (\hat \Xi_2), 
                    \Re (\hat \Xi_3), \Im (\hat \Xi_3))^\top
\]
is the vector of noise processes to which the linear systems 
are subjected; 
hence $\hat \Xi_j~(j=1,2,3)$ satisfies
\[
    d\hat \Xi_j d\hat \Xi_j^\dag = (n+1)dt,~
    d\hat \Xi_j^\dag d\hat \Xi_j = ndt,~
    d\hat \Xi_j^2=d\hat \Xi_j^{2\dag}=0. 
\]
Again, $n>0$ represents the noise strength. 
Finally, $u\in{\mathbb R}^6$ represents the control input for 
the memory. 
Note here it is assumed that both the quadratures of each 
linear system can be controlled. 
In what follows of this section we set $u=0$.

The state transfer can be evaluated in terms of the mean vector and 
the covariance matrix, as discussed in Sec.~II. 
First, the mean vector $\mean{\hat{x}}
=(\mean{\hat{q}_1},\ldots,\mean{\hat{p}_3})^\top$ obeys 
$d\mean{\hat{x}}/dt=A\mean{\hat{x}}-\sqrt{\nu}\beta$, thus 
in the long-time limit we have 
$\mean{\hat x(\infty)}=\sqrt{\nu}A^{-1}\beta
=-2\sqrt{\nu}\beta/(\nu+\Gamma)$. 
That is, the system state becomes Gaussian with mean vector 
parallel to $\beta$, implying that the memory certainly 
acquires information of the input state in the mean sense 
(note that $\nu$ and $\Gamma$ are assumed to be known). 
Next let us consider the covariance matrix of the system: 
\[
    V=\mean{\Delta \hat x \Delta \hat x^\top
                + (\Delta \hat x \Delta \hat x^\top )^\top }/2,~~~
       \Delta \hat x =\hat x-\mean{\hat x}. 
\]
Using the Ito rule, the time evolution of $V$ is found in the 
following form called the Lyapunov differential equation: 
\begin{align}
\label{lyap}
	\dot V = AV + VA^\top
	          + \Gamma(n + 1/2)I_6 
	          + \nu T\Lambda T^\top, 
\end{align}
where
\begin{eqnarray}
& & \hspace*{-2em}
    \Lambda = {\rm diag}
                 \{ \Lambda_1, \Lambda_2, \Lambda_3 \}, 
\nonumber \\ & & \hspace*{-2em}
    \Lambda_j=\begin{pmatrix}
                 N_j + \Re(M_j) + 1/2 & \Im(M_j)  \\
                 \Im(M_j)             & N_j - \Re(M_j) + 1/2
             \end{pmatrix}. 
\nonumber
\end{eqnarray}
Note $\Lambda_2=\Lambda_3$ by assumption. 
Now, to explicitly evaluate the state transfer, let us take 
the source state to be coherent, in which case $M_1=N_1=0$; 
then the steady solution of Eq. \eqref{lyap} is obtained as 
\[
    V_\infty
       =T{\rm diag}\{ V_1, V_2, V_3\}T^\top,~~
    V_j={\rm diag}\{v^+_j, v^-_j\},
\]
where 
\[
   v^\pm_j
	  =\frac{\nu (2 N_j \pm 2 M_j + 1)+\Gamma (1+2n)}
	        {2(\nu + \Gamma )}.
\]
Note $V_2=V_3$. 
Then we have the explicit form of the fidelity between the 
input state $\ket{\psi_{\rm in}}$ and the steady state of the 
memory with its mean appropriately displaced: 
\begin{eqnarray}
\label{uncontrolled fidelity}
& & \hspace*{-2em}
    F=\bra{\psi_{\rm in}}\hat\rho_{\infty}\ket{\psi_{\rm in}}
     =\frac{1}{\sqrt{{\rm det}(V_\infty+V_{\rm in})}}
\nonumber \\ & & \hspace*{-1em}
     =\prod_{\sigma=0,+\mu,-\mu}
       \frac{2(\nu+\Gamma)}
            {2\nu e^\sigma + \Gamma(e^\sigma+1+2n)}, 
\end{eqnarray}
where $V_{\rm in}=T\Lambda T^\top$ is the covariance matrix 
of the input state $\ket{\psi_{\rm in}}$. 
Note that Eq. \eqref{GHZ} is the ideal limit of 
$\ket{\psi_{\rm in}}$. 
Eq. \eqref{uncontrolled fidelity} shows that, when $\Gamma=0$, 
we have $F=1$ without respect to the value of $\mu$. 
This means that the input state is perfectly transferred into 
the memory when it is not coupled to the noisy environment. 
But in reality the system must be subjected to some noise; 
QEC is expected to overcome this issue. 
Motivated by some continuous variable QEC protocols found in the 
literature
\cite{Braunstein1998prl,Braunstein1998nature,van2008note,Aoki2009}, 
let us take the operator $\hat q_i-\hat q_j~(i,j=1,2,3)$ as 
a syndrome. 
Actually, this has desirable properties shown as follows. 
First, it can be seen from the structure of the system dynamics 
\eqref{dynamics} that the syndrome does not contain the unknown 
parameters $\alpha_{\rm in}, N_1$, and $M_1$. 
Second, in the ideal limit $\mu \rightarrow -\infty$, the variance 
of each syndrome is given by 
\begin{equation}
\label{syndrome ensemble}
    \langle \Delta (\hat q_i-\hat q_j)^2 \rangle
        = \frac{\Gamma (2n+1) }{(\nu +\Gamma )}, 
           \quad \forall i\not =j, 
\end{equation}
which takes zero only when the system does not couple to the 
noisy environment. 
This means that, if in the small interval $[t, t+dt)$ only one of 
the three linear systems is subjected to the noise, simultaneous 
``measurement" of the syndrome operators can detect in which 
system that error has occurred and how much it is. 
(The reason why the double quotation is taken is that the syndrome 
operator cannot be directly measured in our setting, but it needs 
to be {\it estimated}; see the next section.) 
Because of these two properties, in the limit 
$\mu\rightarrow -\infty$, the operators $\hat q_i-\hat q_j$ 
certainly plays a role of a syndrome. 
Here it should be noted that $F$ takes the maximum value when 
$\mu=0$; 
that is, without error correction the encoding process merely 
degrades the performance of the state transfer
\footnote{
This can be easily seen in a discrete-variable case; 
For instance, a three-qubit encoded state 
$\ket{\phi}=a\ket{000}+b\ket{111}$ changes due to the first bit 
flip error into $\ket{\phi'_1}=a\ket{100}+b\ket{011}$, hence their 
inner product is $\pro{\phi}{\phi'_1}=0$. 
Likewise, for the second or the third bit-flipping cases the inner 
products are zeros as well. 
Therefore the unconditional state after the error, 
$\hat\rho=(\ket{\phi'_1}\bra{\phi'_1}+\ket{\phi'_2}\bra{\phi'_2}
+\ket{\phi'_3}\bra{\phi'_3})/3$, has no overlap with the encoded 
state, i.e., $F=\bra{\phi}\hat\rho\ket{\phi}=0$. 
That is, without error correction, encoding process can merely 
degrades the input-output fidelity. 
}.
Hence we really need the correction process.

Before closing this section, let us evaluate the entanglement 
of the system state, using essentially the same quantity 
as Eq. \eqref{field_entanglement}. 
Again in the case $N_1=M_1=0$, we have 
\begin{eqnarray}
& & \hspace*{-2em}
     {\cal P}_{\rm sys} 
         := \langle \Delta (\hat q_1-\hat q_2)^2\rangle
                 + \langle \Delta (\hat q_2-\hat q_3)^2\rangle
                    + \langle \Delta (\hat q_3-\hat q_1)^2\rangle
\nonumber \\ & & \hspace*{2em}
     \mbox{}
        + 3\langle \Delta (\hat p_1+\hat p_2+\hat p_3)^2\rangle
\nonumber \\ & & \hspace*{0.3em}
     = \frac{(4.5+3e^\mu)\nu}{\nu+\Gamma}
            + \frac{(7.5+15n)\Gamma}{\nu+\Gamma}. 
\nonumber
\end{eqnarray}
When $\Gamma=0$ and $\mu\rightarrow -\infty$, it is found 
${\cal P}_{\rm sys}\rightarrow 4.5$. 
Hence, together with $\mean{\Delta (\hat q_i-\hat q_j)^2}=0$ in 
this case, the memory state becomes an entangled GHZ-like 
state. 
But in the realistic situation with $\Gamma>0$ such entanglement 
can vanish. 
Particularly in the ideal case $\mu\rightarrow -\infty$ the 
sufficient condition for entanglement, ${\cal P}_{\rm sys}<6$, 
leads to $n<0.1(\nu/\Gamma)-0.1$; 
that is, in the case of thermal noise, 
the averaged photon number of the environment field must 
be less than about one order of magnitude below the S/N rate $\nu/\Gamma$.

%%%%%%%%%%%%%%%%%%%%%%%%%%%%%%%%%%%%%%%%%%%%%%%%%%%%%%%%%%%%%%%%%%%
%%%%%%%%%%%%%%%%%%% Syndrome Filter and Recovering %%%%%%%%%%%%%%%%
%%%%%%%%%%%%%%%%%%%%%%%%%%%%%%%%%%%%%%%%%%%%%%%%%%%%%%%%%%%%%%%%%%%

\section{Syndrome filter and dynamical feedback control}
\label{filtering}

In the previous section it was found that $\hat q_i-\hat q_j$ 
serves as the syndrome through which the error can be detected. 
Since this observable cannot be measured directly in our setting, 
it should be appropriately estimated through indirect measurement. 
In this section we first describe a continuous-time estimator, 
i.e., {\it the filter}, for the syndromes. 
We then present the dynamical feedback controller 
that is based on the syndrome filter.

%%%%%%%%%%%%%%%%%%%%%%%%%%%%%%%%%%%%%%%%%%%%%%%%%%%%%%%%%%%%%%%%%%%

\subsection{Syndrome Filter}

Let us focus on the following two kinds of operator vectors: 
\begin{equation}
\label{syndrome 1}
   \hat s_1 = \tilde B_1\hat x
            = \frac{1}{\sqrt{6}}
              \begin{pmatrix}
                \sqrt{2}(\hat p_1+\hat p_2+\hat p_3)  \\
                \hat q_2 + \hat q_3 - 2\hat q_1  \\
                \sqrt{3}(\hat q_2 - \hat q_3)  \\
              \end{pmatrix}
\end{equation}
and 
\begin{equation}
\label{syndrome 2}
   \hat s_2 = \tilde B_2\hat x
            = \frac{1}{\sqrt{6}}
              \begin{pmatrix}
                \hat q_2 + \hat q_3 - 2\hat q_1  \\
                \sqrt{3}(\hat q_2 - \hat q_3)  \\
              \end{pmatrix}. 
\end{equation}
Here $\tilde B_1:=Z_1T^\top$ and $\tilde B_2:=Z_2T^\top$ are 
isometric matrices with 
\[
   Z_1 = \begin{pmatrix}
           0 & 1 & 0 & 0 & 0 & 0 \\
           0 & 0 & 1 & 0 & 0 & 0 \\
           0 & 0 & 0 & 0 & 1 & 0
         \end{pmatrix},~~
   Z_2 = \begin{pmatrix}
           0 & 0 & 1 & 0 & 0 & 0 \\
           0 & 0 & 0 & 0 & 1 & 0 
         \end{pmatrix}. 
\]
If only one of the three linear systems is subjected to the noise, 
$\hat s_2$ can detect in which system that error 
has occurred. 
Hence, $\hat s_2$ serves as the syndrome operator. 
The first element of $\hat s_1$ is used to 
evaluate entanglement of the memory state.

Our task is to estimate the operator 
$\hat s_1$ or $\hat s_2$ through certain indirect measurement. 
The lower box in Fig. \ref{physical setup} depicts an optical 
configuration that achieves this goal for the case of 
$\hat s_1$; 
that is, we construct the tritter acting on the output fields 
that are reflected at the field-system coupler. 
The observables to be measured by homodyne detectors with 
appropriate LO phase are then given by 
\begin{equation}
\label{output 1}
   d\hat Y_1 = \sqrt{2\nu}\hat s_1 dt 
                +\sqrt{2}(d\hat P_1,~d\hat Q_2,~d\hat Q_3)^\top. 
\end{equation}
Hence measuring $\hat Y_1$ actually implies the indirect 
measurement of $\hat s_1$. 
Note this contains the source fluctuation $d\hat P_1$, hence 
in this case only the mean value of the input state, $\beta$, 
can be kept unknown. 
On the other hand, for the case of $\hat s_2$, two of the three 
output fields are measured; 
\begin{equation}
\label{output 2}
   d\hat Y_2 = \sqrt{2\nu}\hat s_2 dt 
                +\sqrt{2}(d\hat Q_2,~d\hat Q_3)^\top. 
\end{equation}
$\hat Y_2$ does not contain both the mean and covariance of the 
source state, thus we can estimate the syndromes without respect 
to the unknown parameters. 
Note here that the position squeezing of the ancilla fields 
reduces the fluctuation $d\hat Q_2$ and $d\hat Q_3$; 
in particular taking the ideal limit $\mu\rightarrow -\infty$ 
we have $d\hat Y_2 = \sqrt{2\nu}\hat s_2 dt$, thus this is no 
more than the syndrome measurement.

We now describe the filter of $\hat s_i$, which is constructed 
with the measurement results of $\hat Y_i~(i=1, 2)$. 
Let us write Eq. \eqref{output 1} or \eqref{output 2} in the 
following general form:
\begin{equation}
\label{output general}
   d\hat Y = C\hat x dt + D d\hat W. 
\end{equation}
For the case of estimating $\hat s_1$, the matrices correspond to 
$C=\sqrt{2\nu}\tilde B_1=\sqrt{2\nu}Z_1T^\top$ and 
$D=\sqrt{2}(Z_1, O_{3\times 6})$. 
Also for the case of $\hat s_2$, 
$C=\sqrt{2\nu}\tilde B_2=\sqrt{2\nu}Z_2T^\top$ and 
$D=\sqrt{2}(Z_2, O_{2\times 6})$. 
For the simple notation, we do not put the index $i$ on $\hat Y$, 
$C$, and $D$. 
The continuous measurement of $\hat Y$ enables us to construct the 
filter for the dynamics \eqref{dynamics}: 
\begin{eqnarray}
& & \hspace*{-1em}
\label{original filter}
     d\pi(\hat x) 
        = A\pi(\hat x)dt + udt -\sqrt{\nu}\beta dt + K d\bar w, 
\\ & & \hspace*{0.6em}
\label{kalman gain}
     K=(V_{\rm c}C^\top -\sqrt{2\nu}T\Lambda Z^\top )
          (2Z\Lambda Z^\top )^{-1},
\\ & & \hspace*{0.25em}
\label{innovation}
     d\bar w = dy - C\pi(\hat x)dt, 
\end{eqnarray}
where $Z$ implies $Z_1$ or $Z_2$. 
The set of above equations is called the quantum Kalman 
filter 
\cite{Doherty1999,Stockton2004,Edwards2005,Yamamoto2006,
Wiseman2009book}. 
The 6-dimensional c-number vector
$\pi(\hat x_t)={\mathbb E}(\hat x_t|{\cal Y}_t)$ represents 
the quantum conditional expectation of $\hat x_t$ conditioned 
on the set of measurement data ${\cal Y}_t=\{y_s~|~0\leq s\leq t \}$ 
with $y_s$ the measurement result of $\hat Y_s$. 
Note $\pi(\hat x_t)$ is the least mean square estimate of 
$\hat x_t$. 
$V_{\rm c}$ is the conditional covariance matrix 
$V_{\rm c}=\pi( \Delta \hat x \Delta \hat x^\top 
+ (\Delta \hat x \Delta \hat x^\top )^\top )/2$ with 
$\Delta \hat x =\hat x- \pi(\hat x)$, and it obeys the following 
Riccati differential equation: 
\begin{align}
\label{riccati in main context}
    \dot V_{\rm c}
       =&AV_{\rm c}+V_{\rm c}A^\top 
            + \nu T\Lambda T^\top 
              + \Gamma \Big(n+\frac{1}{2}\Big)I_6 \nonumber 
\\
    & -2K Z\Lambda Z^\top K^\top, 
\end{align}
where $K$ is given by Eq. \eqref{kalman gain}. 
Note this is not a stochastic process. 
The filtering equation for the syndrome $\hat s_1$ or $\hat s_2$ 
can be directly obtained from Eq. \eqref{original filter} 
as follows: 
\begin{align}
\label{s_filter}
     d\pi(\hat s) &= d\tilde B\pi(\hat x)
\nonumber \\
     &= \tilde BA\pi(\hat x)dt + \tilde Budt 
           - \sqrt{\nu}\tilde B\beta dt + \tilde BK d\bar w 
\nonumber \\
     &=\tilde A\pi(\hat s)dt + \tilde Budt + \tilde K d\bar w,
\end{align}
where $\tilde A=\tilde BA\tilde B^\top$ and $\tilde K =\tilde BK$. 
Here, $\tilde B$ implies $\tilde B_1$ or $\tilde B_2$. 
Also we have dropped the index of $\hat s_i$ and simply denote $\hat s$. 
The unknown parameter $\alpha_{\rm in}$ does not 
appear in Eq.~\eqref{s_filter} due to $\tilde B\beta=0$. 
Moreover, for the case of estimating $\hat s_2$, the corresponding 
Kalman gain $\tilde{K}=\tilde B_2 K$ does not contain $N_1$ and $M_1$, 
thus the filter can update the estimate of $\hat s_2$ 
without respect to the unknown source state. 
We call Eq. \eqref{s_filter} the {\it syndrome filter}.

%%%%%%%%%%%%%%%%%%%%%%%%%%%%%%%%%%%%%%%%%%%%%%%%%%%%%%%%%%%%%%%%%%%

\subsection{LQG control for the state transfer}

Due to the coupling to the noisy environment, the system state 
must escape from the codespace spanned by $\{ \ket{q,q,q} \}$, but 
this error can be detected by estimating the syndrome operators. 
Therefore, it is expected that a feedback controller minimizing 
the estimated value of the syndromes corrects that error. 
Note that the controller must be a dynamical one so that it 
can deal with the dynamical noise; 
hence, the minimization should be carried out through the whole 
transfer process. 
Fortunately, because of the linearity of both the dynamics and 
the output equation, this requirement is satisfied when employing 
the LQG control strategy 
\cite{Doherty1999,Stockton2004,Edwards2005,Yamamoto2006,
Wiseman2009book}. 
That is, we can construct an optimal control law $u^*$ that 
minimizes the following quadratic-type cost function: 
\begin{align}\label{objective}
    J = \biggl \langle \frac{1}{2}\int^T_0
            ({\hat s}^{\top }Q{\hat s} + u^{\top}Ru)dt 
                \biggr \rangle, 
\end{align}
where $R=R^\top>0$ represents the penalty for the control input. 
Corresponding to the syndrome operators $\hat s_1$ or $\hat s_2$, 
the weighting matrix $Q$ is set to:
\[
    Q_1={\rm diag}\{ 9,3,3\},~~~
    Q_2={\rm diag}\{ 3,3\}. 
\]
It is immediately seen that 
\begin{align}
\label{synsum}
    \hat s^{\top}_1Q_1\hat s_1 
       =& (\hat q_1-\hat q_2)^2
           + (\hat q_2-\hat q_3)^2
              + (\hat q_3-\hat q_1)^2
\nonumber \\
     & +3(\hat p_1+\hat p_2+\hat p_3)^2,
\nonumber \\
    \hat s^{\top}_2Q_2\hat s_2 
       =& (\hat q_1-\hat q_2)^2
           + (\hat q_2-\hat q_3)^2
              + (\hat q_3-\hat q_1)^2. 
\end{align}
Therefore, for both cases the optimal control tries to 
minimize the syndrome errors. 
In the case of $\hat s_1$, the optimal control further takes 
into account the effect of entanglement and tries to keep 
the GHZ-like form of the state. 
The LQG control theory gives the explicit form of the 
(stationary) optimal controller: 
\begin{align}
\label{gain}
       u^* = F\pi(\hat s) = -R^{-1}\tilde B^{\top}P\pi(\hat s),
\end{align}
where $P$ is the solution to the following algebraic Riccati 
equation: 
\begin{align}
    \tilde A^{\top}P + P\tilde A + Q - P\tilde BR^{-1}\tilde B^{\top}P = 0. 
\end{align}
Let us here assume that $R=rI_6>0$. 
Then, corresponding to $\hat s_1$ and $\hat s_2$, the matrix $P$ 
is respectively obtained in the following 
form: 
\begin{align}
    P_1={\rm diag}\{ f_1, f_2, f_2 \},~~~
    P_2={\rm diag}\{ f_2, f_2 \}, 
\end{align}
where
\begin{eqnarray}
& & \hspace*{0em}
    f_1 = -\frac{\nu +\Gamma}{2}
          +\sqrt{\frac{(\nu +\Gamma )^2}{4}+\frac{9}{r}},
\nonumber \\ & & \hspace*{0em}
    f_2 = -\frac{\nu +\Gamma}{2}
          +\sqrt{\frac{(\nu +\Gamma )^2}{4}+\frac{3}{r}}. 
\nonumber
\end{eqnarray}

We here give two brief remarks. 
(i) The first one is about the structure of the optimal 
feedback control input $u^*$. 
In the case of $\hat s_1$, it is given by 
\begin{align}
   u^*=\frac{f_2}{3r}
     \begin{pmatrix}
           \pi(\hat q_2-\hat q_1)+\pi(\hat q_3-\hat q_1) \\
           -f_1\pi(\hat p_1 + \hat p_2 + \hat p_3)/f_2 \\
           \pi(\hat q_1-\hat q_2)+\pi(\hat q_3-\hat q_2) \\
           -f_1\pi(\hat p_1 + \hat p_2 + \hat p_3)/f_2 \\
           \pi(\hat q_1-\hat q_3)+\pi(\hat q_2-\hat q_3) \\
           -f_1\pi(\hat p_1 + \hat p_2 + \hat p_3)/f_2 \\
     \end{pmatrix}. 
\nonumber
\end{align}
The first element can be regarded as a linearization to the 
following sign function ($\lambda=f_2/3r$): 
\[
    \lambda\big[
        {\rm sgn}(\pi(\hat q_2-\hat q_1)) + 
          {\rm sgn}(\pi(\hat q_3-\hat q_1))\big], 
\]
where ${\rm sgn}(x)=+1$ if $x>0$ and $-1$ otherwise. 
Hence, in the small interval $[t,t+dt)$, if the position shift of 
$\hat q_1$ due to the thermal noise is larger than the others, 
the controller adds the largest inverse shift $-2\lambda dt$ on 
$d\hat q_1$ so as to cancel out the error. 
The other elements of $u^*$ have similar meanings. 
It should be maintained that the optimal controller obtained 
in the LQG control setup has this kind of natural structure 
found in the usual QEC scheme. 
(ii) The second remark is related to the no-go theorem 
\cite{niset2009} mentioned in Sec.~I. 
The QEC problem corresponding to our setting is that the 
system's initial state is to be protected by feedback control. 
However, this goal is never accomplished, because, 
in general, a stable Kalman filter forgets the initial 
values of $\pi (\hat x_0)$ and $V_c(0)$; that is, the initial 
Gaussian state with mean $\pi (\hat x_0)$ and covariance $V_c(0)$ 
cannot be protected. 
This is a way to understand the no-go theorem for Gaussian QEC 
from the dynamical control viewpoint.

%%%%%%%%%%%%%%%%%%%%%%%%%%%%%%%%%%%%%%%%%%%%%%%%%%%%%%%%%%%%%%%%%%%%
%%%%%%%%%%%%%%%%%%%%%%%%%%%% Simulation %%%%%%%%%%%%%%%%%%%%%%%%%%%%
%%%%%%%%%%%%%%%%%%%%%%%%%%%%%%%%%%%%%%%%%%%%%%%%%%%%%%%%%%%%%%%%%%%%

\section{Simulation}

In this section, a numerical simulation is provided to demonstrate 
effectiveness of the proposed control scheme. 
As an example of the memory we here take an opto-mechanical 
oscillator shown in FIG.~\ref{fig1}. 
The oscillator must be subjected to thermal noise even at ultra-low 
temperature; 
in this case the noise strength $n$ corresponds to the averaged 
photon number 
$n=({\rm e}^{\hbar\omega_m/k_B T}-1)^{-1}$, where $k_B$, $T$, and 
$\omega_m$ denote the Boltzmann constant, the temperature, and the 
career frequency of the thermal channel, respectively.

The parameters are taken as follows. 
The oscillator couples with the input field with strength 
$\nu/2\pi=30~{\rm kHz}$, while for the thermal channel it is 
$\Gamma/2\pi=1~{\rm Hz}$. 
The thermal channel is with frequecy $\omega_m/2\pi=10~{\rm MHz}$ 
and with temperature $T=4~{\rm K}$, which leads to 
$n=8.8\times 10^3$. 
The mean value of the input state is taken as 
$\alpha_{\rm in}=-230~{\rm Hz}^{1/2}$, 
which leads to $-\sqrt{\nu}\beta =[100, 0, 100, 0, 100, 0]$. 
Note that $|\alpha_{\rm in}|^2$ is the mean photon number per 
unit time.

%%%%%%%%%%%%%%%%%%%%%%%%%%%%%%%%%%%%%%%%%%%%%%%%%%%%%%%%%%%%%%%%%%%%

\subsection{Coherent source state}

First let us consider the case where the source state is taken 
as a coherent state $\ket{\alpha_{\rm in}}$. 
This means that the input fluctuation is set to $N_1=M_1=0$, 
implying that, in addition to the syndrome operators, 
$\hat p_1+\hat p_2+\hat p_3$ can be estimated. 
Hence, we can now apply the feedback control scheme based on 
$\pi(\hat s_1)$.

\begin{figure}[!htbp]
\begin{center}
\includegraphics[width=80mm]{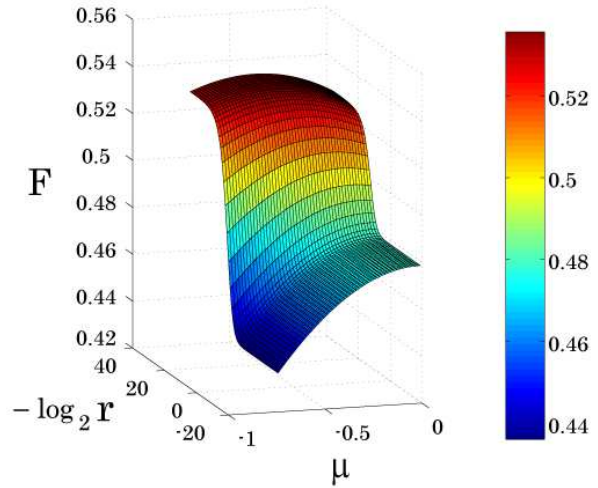}
\caption{
(Color online) 
Fidelity between the input state and the steady state of the 
controlled oscillators, versus the squeezing parameter $\mu$ 
and the control strength $-{\rm log}_2 r$. 
}
\label{fidelity}
\end{center}
\end{figure}

Figure \ref{fidelity} shows how much the encoding and the 
feedback control improve the fidelity between the three-mode 
input state and the oscillators' steady state. 
They are quantified as follows. 
(i) Encoding strength is directly evaluated by the squeezing 
parameter $\mu$; 
actually, as indicated by Eq. \eqref{encoded q-difference}, 
making $-\mu$ bigger means that the input state is going to 
approximate the GHZ-like state \eqref{GHZ}. 
(ii) The LQG optimal control input $u^*$ can take possibly 
a bigger value by tuning the penalty parameter $r$ smaller
(Recall that we have set $R=rI_6$ in Eq.~\eqref{objective}.) 
Hence, $-{\rm log}_2 r$ is interpreted as the control strength. 
(iii) The fidelity is given by the same form as 
Eq. \eqref{uncontrolled fidelity} with $V_\infty$ now 
replaced by the covariance matrix of the ensemble average of 
the {\it controlled} system state, say $V_\infty'$; 
that is, 
$F=1/\sqrt{{\rm det}(V_\infty'+V_{\rm in})}$. 
Note that the controlled system variable $\hat x_t$ stochastically 
changes in time according to the QSDE \eqref{dynamics} with the 
input $u^*$ a function of the estimate $\pi(\hat s_1)$, 
i.e., Eq. \eqref{gain}. 
Hence $V_\infty'$ corresponds to the first $6\times 6$ 
block matrix of the $9\times 9$ covariance matrix of 
$(\hat x^\top, \pi(\hat s_1)^\top)^\top$ that satisfies the 
controlled Lyapunov equation \eqref{augmented lyapunov}; 
see Appendix for the detailed calculation.

Let us now discuss the performance of the LQG optimal feedback 
control. 
It is observed from Fig. \ref{fidelity} that bigger control 
strength indicates bigger fidelity for all value of $\mu$. 
This means that the feedback control can always reduce the 
excess fluctuation brought from the thermal environment, 
without respect to how much the source state is encoded. 
Regarding the encoding strength, however, care should be 
taken in its choice. 
Actually, while larger squeezing enables us to detect a bigger 
error signal which can induce more efficient feedback control, 
at the same time the memory state becomes more sensitive to 
the thermal noise. 
In other words, too much squeezing makes the input state fragile to 
the noise; 
consequently there exists an optimal value of the squeezing 
parameter, and it is about $\mu^*=-0.4$. 
Note this value can be reached within the current technology. 
In this case, the amount of improvement of the fidelity via the feedback 
control compared to the case without encoding and control 
is about $0.05$; but this is not a big improvement. 
One reason to this limitation is that the error taken in this 
paper is the worst one in the sense that each system is 
subjected to the thermal noise for all time; 
this kind of error is not fully 
tractable via the standard QEC protocol. 
Therefore, it is expected that the proposed control scheme 
could show possibly much better performance of the state 
transfer against some weak noise such as simple displacement 
acting only one of the oscillators.

\begin{figure}[!htbp]
\begin{center}
\includegraphics[width=80mm]{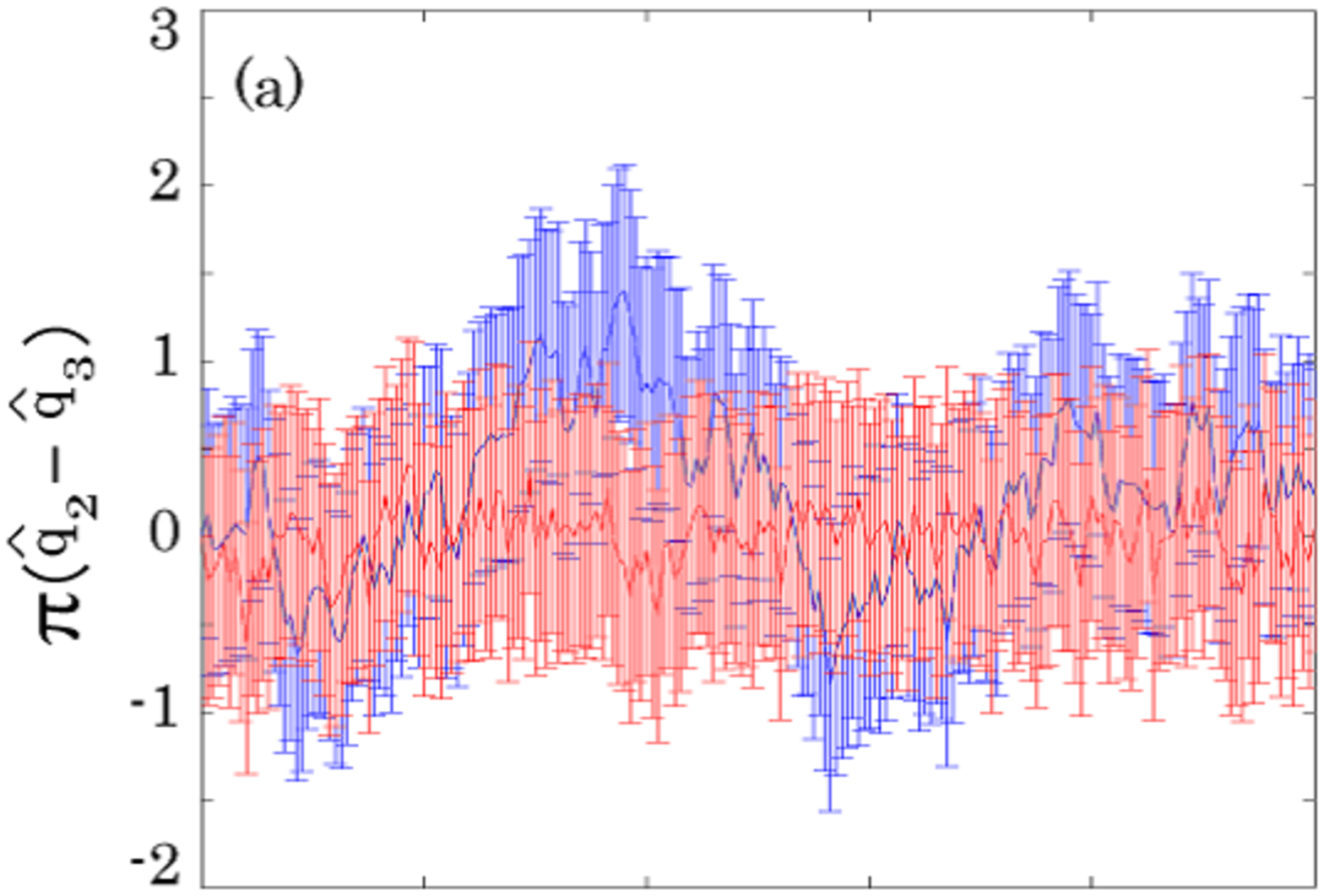}
\includegraphics[width=80mm]{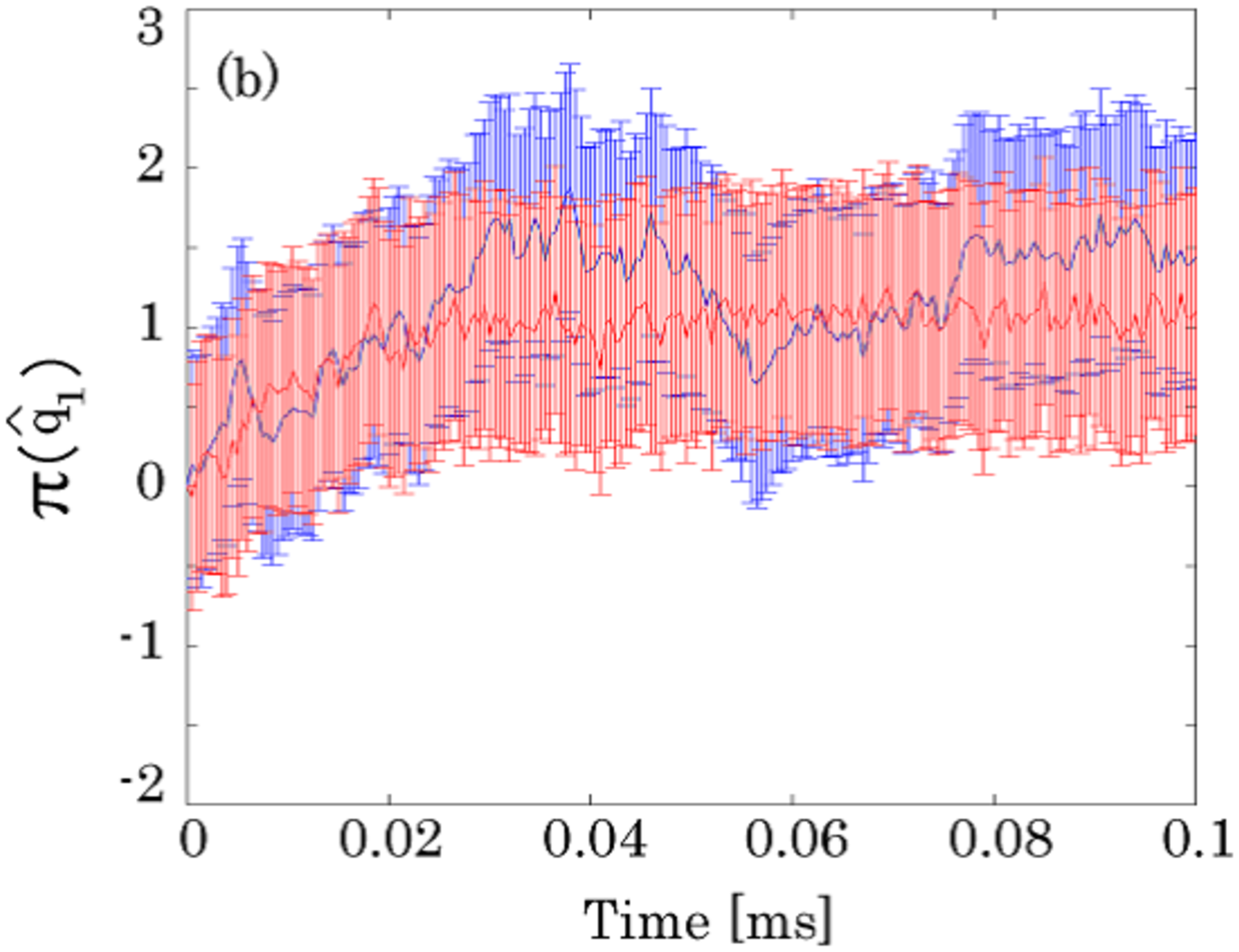}
\caption{
(Color online) 
Time evolutions of (a) $\pi(\hat q_2-\hat q_3)$ and 
(b) $\pi(\hat q_1)$. 
For both plots, the less (red) and larger (blue) fluctuating 
lines correspond to the cases where the optimal feedback 
control is performed or not, respectively. 
}
\label{time evolution}
\end{center}
\end{figure}

Next, Fig. \ref{time evolution} shows the time evolutions of 
(a) the syndrome $\pi(\hat q_2-\hat q_3)$ and 
(b) the first oscillator's position $\pi(\hat q_1)$. 
For both plots, the less (red) and larger (blue) fluctuating 
lines correspond to the cases where the optimal feedback control 
is performed or not, respectively. 
The error bars show the standard deviation of the estimation 
error, which is calculated from the solution to the Riccati equation 
\eqref{riccati in main context}. 
The parameters are taken as $\mu=-0.4$ and $r=10^{-9}$, which 
were shown in Fig. \ref{fidelity} to be the optimal values 
attaining the maximum fidelity. 
Note that, because $\alpha_{\rm in}$ is unknown, 
$\hat q_1$ cannot be exactly estimated; but we here 
plot the exact value of $\pi(\hat q_1)$ just for demonstration. 
Figure \ref{time evolution} (a) demonstrates that the fluctuation 
of the syndrome is certainly suppressed by the feedback control; 
this is the result that should be expected, since the controller 
is designed to minimize the syndrome errors. 
On the other hand, it is not straightforwardly expected that 
controlling $\hat q_1$ would actually work well, but Fig. 
\ref{time evolution} (b) demonstrates that the feedback 
control based on the syndrome estimation is fairly effective 
for correcting the error acting on each position of the 
oscillator.

%%%%%%%%%%%%%%%%%%%%%%%%%%%%%%%%%%%%%%%%%%%%%%%%%%%%%%%%%%%%%%%%%%%%

\subsection{Squeezed source state}

Here we are concerned with the situation where the source 
state is squeezed. 
The squeezing parameter is taken as 
$\mu_1={\rm log}(2M_1+2N_1+1)$ with $M_1$ real, and $N_1$ and 
$M_1$ are both unknown in addition to $\alpha_{\rm in}$. 
We then have to use the filter for $\hat s_2$, which does not 
contain $N_1$ and $M_1$. 
But for comparison we further consider a squeezed source state 
with known covariance and unknown mean value; 
in this case the Kalman filter for $\hat s_1$ can be used for 
feedback control. 
This comparison will reveal how much the additional information 
(i.e., the estimate of $\hat p_1+\hat p_2+\hat p_3$) improves 
the control performance at the expense of limiting the class 
of input states.

\begin{figure}[htbp]
\begin{center}
\includegraphics[width=80mm]{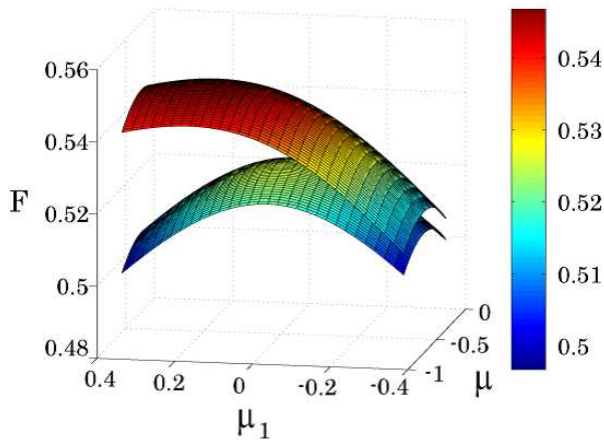}
\caption{
(Color online) 
Fidelity between the input state and the controlled 
oscillators' steady state, versus the squeezing parameters 
$\mu$ and $\mu_1$. 
The upper surface corresponds to the case where the estimate 
$\pi(\hat s_1)$ is used to control, while the lower surface 
does the case with $\pi(\hat s_2)$ only available for estimation 
and control. 
}
\label{fidelity comparison}
\end{center}
\end{figure}

Figure \ref{fidelity comparison} shows the fidelity between 
the input state and the controlled oscillators' steady state, 
versus the squeezing parameters $\mu$ and $\mu_1$. 
The control penalty is $r=10^{-9}$. 
In the figure the upper and lower surfaces correspond to the 
filter for $\hat s_1$ and $\hat s_2$, respectively. 
First, a notable fact observed from the lower surface is that, 
when aiming to transfer the input state with the completely 
unknown source state, the maximum fidelity is attained when 
$\mu_1=0$; 
that is, squeezing the source state always decreases the 
fidelity. 
This is consistent with the standard understanding that an 
unknown squeezed state is in general fragile to the thermal noise, 
because it randomly rotates the phase of the state, 
which as a result brings that state into a mixed state. 
On the other hand, the upper surface in the figure shows that, 
when transferring the input state with known fluctuation, which 
means that $\hat s_1$ can be used for estimation and control, 
squeezing the source state improves the fidelity. 
Note that the improvement is observed if $\mu_1>0$ and $\mu<0$; 
that is, the source state is now momentum-squeezed while the 
ancilla states are position-squeezed. 
Actually in this case the input state more closely approximates 
the GHZ-like state, which is an eigenstate of 
$\hat p_1+\hat p_2+\hat p_3$. 
Therefore, the additional information 
$\pi (\hat p_1+\hat p_2+\hat p_3 )$ should certainly improves 
the control performance of the state transfer. 
However, as in the previous case, too much squeezing degrades 
the fidelity, hence $\mu_1$ must be optimized. 

%%%%%%%%%%%%%%%%%%%%%%%%%%%%%%%%%%%%%%%%%%%%%%%%%%%%%%%%%%%%%%%%%%%%
%%%%%%%%%%%%%%%%%%%%%%%%%%%% Conclusion %%%%%%%%%%%%%%%%%%%%%%%%%%%%
%%%%%%%%%%%%%%%%%%%%%%%%%%%%%%%%%%%%%%%%%%%%%%%%%%%%%%%%%%%%%%%%%%%%

\section{Conclusion}

In this paper we proposed a new feedback control scheme for 
state transfer, which has the form of QEC. 
In particular, due to the Gaussianity of the input state 
and the linearity of the memory dynamics, the celebrated 
Kalman filtering and LQG feedback control were employed. 
We have considered a specific system, but the proposed control 
strategy is applicable to any linear quantum system, as 
long as the output process can be properly defined. 
Although it was shown in the example that the controller is 
not a very effective one, it is expected that it can show 
much bigger improvement for some linear systems having 
relatively weak noise. 
Also possible combination with a feedforward control scheme 
\cite{Filip2009,Filip2010} could be helpful to attain better 
performance for Gaussian state transfer.

%%%%%%%%%%%%%%%%%%%%%%%%%%%%%%%%%%%%%%%%%%%%%%%%%%%%%%%%%%%%%%%%%%%%
%%%%%%%%%%%%%%%%%%%%%%%%%%%% Appendix %%%%%%%%%%%%%%%%%%%%%%%%%%%%%%
%%%%%%%%%%%%%%%%%%%%%%%%%%%%%%%%%%%%%%%%%%%%%%%%%%%%%%%%%%%%%%%%%%%%

\appendix

\section{Covariance matrix of the controlled memory state}

The whole closed-loop system of the controlled memory 
and the estimator is subjected to the following QSDE: 
\begin{align*}
     d\hat z &= 
         \begin{pmatrix}
             A & F \\
             \tilde KC & \tilde A-\tilde K C \tilde{B}^\top + \tilde{B}F
         \end{pmatrix}\hat z dt
         +\begin{pmatrix}
               B \\
               \tilde K D
             \end{pmatrix}d\hat W, 
\end{align*}
where $\hat z =(\hat x^\top, \pi(\hat s)^\top)^\top$. 
Then the covariance matrix 
$V_z=\langle \Delta\hat z\Delta\hat z^\top + 
(\Delta\hat z\Delta\hat z^\top)^\top\rangle /2$ with 
$\Delta\hat z=\hat z-\langle\hat z \rangle$ changes in time 
according to the following Lyapunov equation: 
\begin{widetext}
\begin{align}
\label{augmented lyapunov}
    &\frac{d}{dt}V_z=
         \begin{pmatrix}
             A & F \\
             \tilde KC & \tilde A-\tilde K C\tilde{B}^\top +\tilde{B}F
         \end{pmatrix}V_z
      +V_z
         \begin{pmatrix}
             A & F \\
             \tilde KC & \tilde A-\tilde K C\tilde{B}^\top +\tilde{B}F
         \end{pmatrix}^\top 
      +
         \begin{pmatrix}
             B \\
             \tilde K D
         \end{pmatrix}
         \begin{pmatrix}
             \Lambda & \\
             & (n_T+1/2)I_6
         \end{pmatrix}
         \begin{pmatrix}
             B^\top, & D^\top \tilde K ^\top 
         \end{pmatrix}. 
\end{align}
Now, the covariance matrix of the controlled memory state, $V'$, 
corresponds to the first $6\times 6$ block matrix of $V_z$. 
In this case, the steady solution $V'_\infty$ can be explicitly 
obtained as follows: 
\begin{align*}
      V'_\infty
        =&-\frac{1}{2}(2A-K C+F\tilde{B})^{-1}
           \biggl\{
               \begin{pmatrix}
                  A-K C+F\tilde{B}, & -F\tilde{B}
               \end{pmatrix}
               \begin{pmatrix}
                  B \\
                  \tilde K D
               \end{pmatrix}
               \begin{pmatrix}
                  \Lambda & \\
                  & (n_T+1/2)I_6
               \end{pmatrix}
               \begin{pmatrix}
                  B^\top, & D^\top \tilde K^\top 
               \end{pmatrix}
\nonumber \\
       &\times
            \begin{pmatrix}
                A^\top -C^\top K ^\top +\tilde{B}^\top F^\top \\
                -\tilde{B}^\top F^\top 
            \end{pmatrix}
            (A-K C)^{-1}(F\tilde{B}+A)^{-1}
          +\Gamma(n_{\rm T} + 1/2)I_6 + \nu T\Lambda T^\top \biggr\}. 
\label{vxx}
\end{align*}
\end{widetext}
Note that as $r\rightarrow 0$ (i.e., cheap control), we have 
$V'_\infty \rightarrow V_{\rm c}(\infty)$, which is the steady 
solution to Eq. \eqref{riccati in main context}. 
That is, in this limit, the quantum fluctuation of the controlled 
memory state is maximally reduced down to the level of 
estimation error.

%%%%%%%%%%%%%%%%%%%%%%%%%%%%%%%%%%%%%%%%%%%%%%%%%%%%%%%%%%%%%%%%%%%%
%%%%%%%%%%%%%%%%%%%%%%%%%%%% References %%%%%%%%%%%%%%%%%%%%%%%%%%%%
%%%%%%%%%%%%%%%%%%%%%%%%%%%%%%%%%%%%%%%%%%%%%%%%%%%%%%%%%%%%%%%%%%%%

%

\end{document}